\documentclass[%
 superscriptaddress,
 onecolumn,
 nofootinbib,
  amsmath,amssymb,
  aps,
 pra,
  longbibliography,
  floatfix,
 ]{revtex4-1}

\usepackage[normalem]{ulem}
\usepackage{soul}
\usepackage{hyperref}
\usepackage{amsmath}
\usepackage{amssymb}
\usepackage{amsthm}
\usepackage{graphicx}

\RequirePackage{times}
\RequirePackage{mathptm}

\usepackage{url}
\usepackage{yfonts}
\usepackage{color}
\usepackage{manfnt}

\theoremstyle{definition}

\newcommand{\x}{\mathbf{x}}

\newcommand{\ket}[1]{\left| #1 \right>}

\newcommand{\iprod}[2]{\langle #1 | #2 \rangle}

\newcommand{\oprod}[2]{| #1 \rangle\langle #2 |}

\begin{document}
	
\title{A Non-Probabilistic Model of Relativised Predictability in Physics}
	
\author{Alastair A. Abbott}
\email{a.abbott@auckland.ac.nz}
\homepage{http://www.cs.auckland.ac.nz/~aabb009}

\affiliation{Department of Computer Science, University of Auckland,
Private Bag 92019, Auckland, New Zealand}
\affiliation{Centre Cavaill\`es, R\'epublique des Savoirs, USR 3608, CNRS, Coll\`ege de France and \'Ecole Normale Sup\'erieure, 29 rue d'Ulm, 75005 Paris, France}

\author{Cristian S. Calude}
\email{cristian@cs.auckland.ac.nz}
\homepage{http://www.cs.auckland.ac.nz/~cristian}

\affiliation{Department of Computer Science, University of Auckland,
Private Bag 92019, Auckland, New Zealand}

\author{Karl Svozil}
\email{svozil@tuwien.ac.at}
\homepage{http://tph.tuwien.ac.at/~svozil}

\affiliation{Institute for Theoretical Physics,
Vienna  University of Technology,
Wiedner Hauptstrasse 8-10/136,
1040 Vienna,  Austria}

\affiliation{Department of Computer Science, University of Auckland,
Private Bag 92019, Auckland, New Zealand}

\date{\today}

\begin{abstract}
%
Little effort has been devoted to studying generalised notions or models of (un)predictability, yet is an important concept throughout physics and plays a central role in quantum information theory, where key results rely on the supposed inherent unpredictability of measurement outcomes.
In this paper we continue the programme started in~\cite{DBLP:conf/birthday/AbbottCS15} developing a general, non-probabilistic model of (un)predictability in physics.
We present a more refined model that is capable of studying different degrees of ``relativised'' unpredictability.
This model is based on the ability for an agent, acting via uniform, effective means, to predict correctly and reproducibly the outcome of an experiment using finite information extracted from the environment.
We use this model to study further the degree of unpredictability certified by different quantum phenomena, showing that quantum complementarity guarantees a form of relativised unpredictability that is weaker than that guaranteed by Kochen-Specker-type value indefiniteness.
We exemplify further the difference between certification by complementarity and value indefiniteness by showing that, unlike value indefiniteness, complementarity is compatible with the production of computable sequences of bits.
\end{abstract}


\maketitle

\section{Introduction}

Many physical processes and phenomena are intuitively thought of as unpredictable: the roll of a die, the evolution of weather systems, and the outcomes of quantum measurements, to mention a few.
While \emph{ad hoc} definitions of unpredictability may exist within certain domains, little work has been done towards developing a more general understanding of the concept.
Although domain specific notions of unpredictability may help describe and categorise phenomena within the domain, the concept of unpredictability has a much more central and important role in quantum information theory.

Many of the advantages promised by quantum information theory and cryptography rely critically on the belief that the outcomes of quantum measurements are intrinsically unpredictable~\cite{zeil-05_nature_ofQuantum,Fitzsimons:2013kk}.
This belief underlies the use of quantum random number generators to produce ``quantum random'' sequences that are truly unpredictable (unlike pseudo-randomness)~\cite{stefanov-2000} and the generation of cryptographic keys unpredictable to any adversary~\cite{Fitzsimons:2013kk}.
Such claims of quantum unpredictability are generally based on deeper theoretical results---such as the Kochen-Specker~\cite{kochen1} and Bell~\cite{bell-66} theorems, or quantum complementarity---but nonetheless remain informal intuition.

The quantum cryptography community has used a probability theoretic approach to try and make use of, and quantify the degree of unpredictability in quantum information theoretical situations, in particular by following the cryptographic paradigm of adversaries with limited side-information~\cite{Berta:2010vn}.
This approach, while suitable in such cryptographic situations precisely because of its epistemic nature~\cite{dynes:031109}, relies on the probabilistic formalism of quantum mechanics and the subsequently assumed unpredictability.
In order to fully understand and study the degree of quantum unpredictability and randomness, it is instead crucial to have more general models of unpredictability to apply.

Historically, little work has been devoted to such generalised notions of unpredictability.
In~\cite{DBLP:conf/birthday/AbbottCS15} we discussed in some detail the most notable approaches, in particular those of Popper~\cite{popper-50i}, Wolpert~\cite{Wolpert:2008aa}, and Eagle~\cite{Eagle:2005ys}.
In response to these approaches, we outlined a new model based around the ability for a predicting agent, acting via uniform, effective means, to predict correctly and reproducibly the outcome of an experiment using some finite information the agent extracts from the ``environment'' as input.


This model allowed us to consider a specific, ontic, form of unpredictability which was particularly suited for analysing the type of unpredictability quantum mechanics claims to provide.
However, this strong form of unpredictability is too strong in many cases and failed to capture the possible different degrees of unpredictability: what is predictable for one agent may not be for another with different capabilities.

In this paper we refine and improve this model of (un)predictability, providing a more nuanced, relativised notion of unpredictability that can take into account the epistemic limits of an observer, something crucial, for example, in chaotic systems~\cite{Werndl:2009nx}.
This also provides the ability to look at the degree of unpredictability guaranteed by different possible origins of quantum unpredictability.
We examine one such case---that of quantum complementarity---in detail, and show that it provides a weaker form of unpredictability than that arising from Kochen-Specker-type value indefiniteness as discussed in~\cite{DBLP:conf/birthday/AbbottCS15}


\section{Relativised model of predictability}

The model of (un)predictability that we proposed in~\cite{DBLP:conf/birthday/AbbottCS15} is based around the ability of an agent to, in principle, predict the outcome of a physical experiment.
By using computability theory---motivated by the Church-Turing thesis---to provide a universal framework in which prediction can occur, this information-theoretical approach allows different physical systems and theories to be uniformly analysed.

Here we refine and extend this model to be able to relativise it with respect to the means/resources  of the predicting agent.
This gives our model an epistemic element, where our previous and more objective model can be obtained as the limit case.
In this framework we can consider the predictive capabilities of an agent with limited capacities imposed by practical limitations, or under the constraints of physical hypotheses restricting such abilities.

Before we proceed to present our model in detail, we will briefly outline the key elements comprising it.
\begin{enumerate}
	\item The specification of an experiment $E$ for which the outcome must be predicted.
	\item A predicting agent or ``predictor'', which must predict the outcome of the experiment.
	We model this as an effectively computable function, a choice which we will justify further.
	\item An extractor $\xi$ is a physical device the agent uses to (uniformly) extract information pertinent to prediction that may be outside the scope of the experimental specification $E$. This could be, for example, the time, measurement of some parameter, iteration of the experiment, etc.
	\item A prediction made by the agent with access to a set $\Xi$ of extractors. The set of extractors $\Xi$ provides the relativisation of the model.
\end{enumerate}

This model is explicitly a non-probabilistic one, a fact that may seem overly restrictive given that highly probable events seem predictable.
However, the uncertainty present in ``high probabilities'' represents an important latent unpredictability in such processes, and certainty is needed if predictions are to be related to definite properties of physical systems~\cite{epr}, as 
in quantum scenarios, for example.

It should be noted that our model does not assess the ability to make statistical predictions about physical processes (as one might about the throw of a dice, for example)---as probabilistic models might---but rather the ability to predict precise measurement outcomes.

We will next elaborate on the individual aspects of the model.

\subsection{Predictability model}

\textbf{Experimental specification.}
An experiment is a finite specification for which the outcome is to be predicted.
We restrict ourselves to the case where the result of the experiment, i.e.\ the value to be predicted, is a single bit: 0 or 1.
However, this can readily be generalised for any finite outcome.
On the other hand it does not make sense to predict an outcome requiring an infinite description, such as a real number, since this can never be measured exactly.
In such a case the outcome would be an approximation of the real---a rational number, and thus finitely specifiable.

The experimental specification, being finite, cannot normally specify exactly the required setup of the experiment, as a precise description of experimental conditions generally involves real-valued parameters.
Rather, it is expressed with finite precision 
by the experimenter within their limited capacities---making use, for example, of the pertinent symmetries to describe the experiment.
A particular trial of $E$ is associated with the parameter $\lambda$ which fully describes the ``state of the universe'' in which the trial is run.
As an example, one could consider $E$ to specify the flipping of a certain coin, or it could go further and specify, up to a certain accuracy, the initial dynamical conditions of the coin flip.
In both cases, $\lambda$ contains further details---such as the exact initial conditions---which could be used by an agent in trying to predict the result of $E$.

The parameter $\lambda$ will generally%
\footnote{If one insists on a discrete or computational universe---whether it be as a ``toy'' universe, in reality or in virtual reality---then $\lambda$ could be conceived as a finite quantity. This is, however, the exception, and in the orthodox view of real physical experiments $\lambda$ would be infinite, even if the prediction itself is discrete or finite, so we will adopt this view here.}
be ``an infinite quantity''---for example, an infinite sequence or a real number---structured in an unknown manner.
Forcing a specific encoding upon $\lambda$, such as a real number, may impose an inadequate structure (e.g.\ metric, topological)  which is not needed for what follows.
While $\lambda$ is generally not in its entirety an obtainable quantity, it contains any information that may be pertinent to prediction---such as the time at which the experiment takes place, the precise initial state, any hidden parameters, etc.---and any predictor can have practical access to a finite amount of this information.
We can view $\lambda$ as a resource from which one can extract finite information in order to try and predict the outcome of the experiment $E$.

\textbf{Predicting agent.}
The predicting agent (or ``predictor'') is, as one might expect, the agent trying to predict the outcome of a particular experiment, using potentially some data obtained from the system (i.e.\  from $\lambda$) to help in the process.
Since such an agent should be able to produce a prediction in a finite amount of time via some uniform procedure, we need the prediction to be \emph{effective}.

Formally, we describe a predicting agent as a computable function $P_E$ (i.e.\ an algorithm) which halts on every input and outputs either 0,1, or ``prediction withheld''.
Thus, the agent may refrain from making a prediction in some cases if it is not certain of the outcome.
$P_E$ will generally be dependent on $E$, but its definition as an abstract algorithm means \emph{it must be able to operate without interacting  with the subsystem whose behaviour it predicts}.
This is necessary to avoid the possibility that the predictor affects the very outcome it is trying to predict.

We note finally that the choice of computability as the level of effectivity required can be strengthened or weakened, as long as some effectivity is kept. Our choice of computability is motivated by the Church-Turing thesis, a rather robust assumption~\cite{sep-church-turing}.

\textbf{Extractor.}
An extractor is a physically realisable device which a predicting agent can use to extract (finite) useful data that may not be a part of the description of $E$ from $\lambda$ to use for prediction---i.e.\ as input to $P_E$.
In many cases this can be viewed as a measurement instrument, whether it be a ruler, a clock, or a more complicated device.

Formally, an extractor 
produces  a finite string of bits $\xi(\lambda)$ 
which can be physically realised without altering the system, i.e.\ passively.
In order to be used by $P_E$ for prediction, $\xi(\lambda)$ should be finite and effectively codable, e.g.\ as a binary string or a rational number.

\textbf{Prediction.}
We define now the notion of a correct prediction for a predicting agent having access to a fixed (finite or infinite) set $\Xi$ of extractors.

Given a particular extractor $\xi$, we say the prediction of a run of $E$ with parameter $\lambda$ is \emph{correct for $\xi$} if the output $P_E(\xi(\lambda))$ is the same as the outcome of the experiment.
That is, it correctly predicts $E$ when using information extracted from $\lambda$ by $\xi$ as input.

However, this is not enough to give us a robust definition of predictability, since for any given run it could be that we predict correctly by chance.
To overcome this possibility, we need to consider the behaviour of repeated runs of predictions.

A \emph{repetition procedure for $E$} is an algorithmic procedure for resetting and repeating the experiment $E$.
Generally this will be of the form ``$E$ is prepared, performed and reset in a specific fashion''.
The specific procedure is of little importance, but the requirement is needed to ensure the way the experiment is repeated cannot give a predicting agent power that should be beyond their capabilities or introduce mathematical loopholes by ``encoding'' the answer in the repetitions;
both the prediction and repetition should be performed algorithmically.

We say the predictor $P_E$ is correct for $\xi$ if for any $k$ and any repetition procedure for $E$ (giving parameters $\lambda_1, \lambda_2, \dots$ when $E$ is repeated) there exists an $n\ge k$ such that after $n$ repetitions of $E$ producing the outputs $x_1,\dots,x_n$, the sequence of predictions $P_E(\xi(\lambda_1)),\dots,P_E(\xi(\lambda_n))$:
\begin{enumerate}
	\item contains $k$ correct predictions,
	\item contains no incorrect prediction; e.g. the remaining $n-k$ predictions are withheld.
\end{enumerate}

From this notion of correctness we can define predictability both relative to a set of extractors, and in a more absolute form.

Let $\Xi$ be a set of extractors.
An experiment $E$ is \emph{predictable for $\Xi$} if there exists a predictor $P_E$ and an extractor $\xi\in\Xi$ such that $P_E$ is correct for $\xi$.
Otherwise, it is \emph{unpredictable for $\Xi$}.

This means that $P_E$ has access to an extractor $\xi\in\Xi$ which, when using this extractor to provide input to $P_E$, can be made to give arbitrarily many correct predictions by repeating $E$ enough (but finitely many) times, without ever giving an incorrect prediction.

The more objective notion proposed in~\cite{DBLP:conf/birthday/AbbottCS15} can be recovered by considering all possible extractors.
Specifically, an experiment is \emph{(simply) predictable} if there exists a predictor $P_E$ and an extractor $\xi$ such that $P_E$ is correct for $\xi$.
Otherwise, it is (simply) unpredictable.

The outcome $x$ of an \emph{single trial} of the experiment $E$ performed with parameter $\lambda$ is \emph{predictable (for $\Xi$)} if $E$ is predictable (for $\Xi$) and $P_E(\xi(\lambda))=x$.
Otherwise, it is unpredictable (for $\Xi$). 
We emphasise here that the predictability of the result of a single trial is predictability \emph{with certainty}.

\subsection{Relativisation}

While the notion of simple predictability provides a very strong notion of unpredictability---one that seems to correspond to what is often meant in the context of quantum measurements~\cite{DBLP:conf/birthday/AbbottCS15}---in some physical situations, particularly in classical physics, our inability to predict would seem to be linked to our epistemic lack of information, often due to measurement.
Put differently, unpredictability is a result of only having access to a set $\Xi$ of extractors of limited power.
Our relativised model of prediction attempts to capture this, defining predictability relative to a given set of extractors $\Xi$.

\subsubsection{Specifying the set of extractors $\Xi$}

In defining this notion, we deliberately avoided saying anything about how $\Xi$ should be specified.
Here we  outline two possible ways this can be done.

The simplest, but most restrictive, way would be to explicitly specify the set of extractors.
As an example, let us consider the experiment of firing a cannonball from a cannon and the task of predicting where it will land (assume for now that the muzzle velocity is known and independent of firing angle).
Clearly, the position will depend on the angle the cannonball is fired at.
Then, if we are limited to measuring this with a ruler, we can consider, for example, the set of extractors

$$\Xi=\{\xi \mid \text{$\xi(\lambda)=(x,y)$ where $x$ and $y$ are the muzzle position to an accuracy of 1cm}\}$$

and then consider predictability with respect to this set $\Xi$.
(For example, by using trigonometry to calculate the angle of firing, and then where the cannonball will land.)

Often it is unrealistic to characterise completely the set of extractors available to an agent in this way---think about a standard laboratory full of measuring devices that can be used in various ways.
Furthermore, such devices might be able to measure properties indirectly, so we might not be able to characterise the set $\Xi$ so naively.
Nonetheless, this can allow simple consideration and analysis of predictability in various situations, such as under-sensitivity to initial conditions.

A more general approach, although often requiring further assumptions, is to limit the ``information content'' of extractors.
This avoids the difficulty of having to explicitly specify $\Xi$.
Continuing with the same example as before, we could require that no extractor $\xi\in\Xi$ can allow us to know the firing angle better than $1^\circ$.
This circumvents any problems raised by the possibility of indirect measurement, but of course requires us to have faith in the assumption that this is indeed the case;
it could be possible that we \emph{can} extract the angle better than this, but we simply don't know how to do it with our equipment.
(This would not be a first in science!)
Nonetheless, this approach captures well the epistemic position of the predicting agent.

Let us formalise this more rigorously.
We hypothesise that we cannot do any better than a hypothetical extractor $\xi'$ extracting the desired physical quantity.
Then we characterise $\Xi$ by asserting: for all $\xi\in\Xi$ there is no computable function $f$ such that for every parameter $\lambda$, $f(\xi(\lambda))$ is more accurate than $\xi'$.
Obviously, the evaluation of ``more accurate'' requires a (computable) metric on the physical quantity extracted, something not unreasonable physically given that observables tend to be measured as rational numbers as approximations of reals~\cite{Longo:2010pt}.

This general approach would need to be applied on a case by case basis, given assumptions about the capabilities available to the predicting agent.
Assumptions have to be carefully justified and, ideally, subject themselves to experimental verification.

Either of these approaches, and perhaps others, can be used with our relativised model of prediction.
In any such case of relativisation, one would need to argue that the set $\Xi$ unpredictability is proven for is relevant physically.
This is unavoidable for any epistemic model of prediction.

\subsubsection{A detailed example}

Let us illustrate the use of relativised unpredictability with a more interesting example of an experiment which is  predictable, but its intuitive unpredictability is well captured by the notion of relativised unpredictability.
In particular, let us consider a simple chaotic dynamical system.
Chaos is often considered to be a form of unpredictability, and is characterised by sensitivity to initial conditions and the mixing of nearby dynamical trajectories~\cite{Werndl:2009nx}.
However, chaos is, formally, an asymptotic property~\cite{Paul:2009fv}, and we will see that as a result the unpredictability of chaotic systems is not so simple as might be initially suspected.

For simplicity, we will take the example of the dyadic map, i.e. the operation on infinite sequences defined by $d(x_1x_2x_3\dots) = x_2x_3\dots$, as in~\cite{DBLP:conf/birthday/AbbottCS15}.
We work with this example since it is mathematically clear and simple, and is an archetypical example of a chaotic system, being topologically conjugate to many other well-known systems~\cite{Devaney-1989}.
However, the analysis could equally apply to more familiar (continuous) chaotic physical dynamics, such as that of a double pendulum.

Let us consider the hypothetical experiment $E_k$  (for fixed $k\ge 1$) which  involves iterating the dyadic map $k$ times (i.e.\ $d^k$) on an arbitrary ``seed'' $\x=x_1x_2\dots$.
The outcome of the experiment is then taken to be the first bit of the resulting sequence $d^k(\x)=x_{k+1}x_{k+2}\dots$, i.e.\ $x_{k+1}$.
This corresponds to letting the system evolve for some fixed time $k$ before measuring the result.

While the shift $d$ (and hence $d^k$) is chaotic and generally considered to be unpredictable, it is clearly simply predictable if we have an extractor that can ``see'' (or measure) more than $k$ bits of the seed.
That is, take the extractor $\xi_k(\lambda_\x) = x_{k+1}$ which clearly extracts only finite information, and the identity Turing machine $I$ as $P_{E_k}$ so that, for any trial of $E_k$ with parameter $\lambda_\x$ we have $P_{E_k}(\xi_k(\lambda_x)) = I(x_{k+1})=x_{k+1}$, which is precisely the result of the experiment.

On the other hand, if we consider that there is some limit $l$ on the ``precision''  of measurement of $\x$ that we can perform, the experiment is unpredictable relative to this limited set of extractors $\Xi_l$ defined  such that for every sequence $\x$ and every computable function $f$  there exists $\lambda$ such that for all
$j>l$, $f(\xi(\lambda))\not=x_j$. It is clear that for $l=k$, \emph{given the limited precision of measurements assumption}, 
the experiment $E_k$ is unpredictable for $\Xi_k$. Indeed,   if this were not the case, the pair $(\xi,P_{E_k})$ allowing prediction would make arbitrarily many correct predictions, thus  contradicting the assumption on limited precision of measurements.

This example may appear somewhat artificial, but this is not necessarily so.
If one considers the more physical example of a double pendulum, as mentioned earlier, one can let  it evolve for a fixed time $t$ and attempt to predict its final position (e.g. above or below the horizontal plane) given a set limit $l$ on the precision of any measurement of the initial position in phase space.
If the time $t$ is very short, we may well succeed, but for long $t$ this becomes unpredictable.

This re-emphasises that chaos is an asymptotic property, occurring only strictly at infinite time.
While in the limit it indeed seems to correspond well to unpredictability, in finite time the unpredictability of chaotic systems is relative: a result of our limits on measurement.
Of course, in physical situations such limits may be rather fundamental: thermal fluctuation or quantum uncertainty seem to pose very real limits on measurement precision~\cite{Longo:2010pt}, although in most situations the limits actually obtained are of a far more practical origin.

\section{Unpredictability in quantum mechanics}

As we discussed in the introduction, the outcomes of individual quantum measurements are generally regarded as being inherently unpredictable, a fact that plays an important practical role in quantum information theory~\cite{10.1038/nature09008,gisin-qc-rmp}.
This unpredictability has many potential origins, of which quantum value indefiniteness is perhaps one of the most promising candidates to be used to certify it more formally.

\subsection{Quantum value indefiniteness}
\label{sec:VI}

Value indefiniteness is the notion that the outcomes of quantum measurements are not predetermined by any function of the observables and their measurement contexts---that there are no hidden variables.
It is thus a formalised notion of indeterminism, and the measurement of such observables results in an outcome not determined before the measurement took place.

While it is possible to hypothesise value indefiniteness in quantum mechanics~\cite{zeil-99}, its importance comes from the fact that it can be proven (for systems represented in dimension three or higher Hilbert space) to be true  under simple classical hypotheses via the Kochen-Specker theorem~\cite{kochen1,2012-incomput-proofsCJ,2015-AnalyticKS}.
We will not present the formalism of the Kochen-Specker theorem here, but just  emphasise that this gives value indefiniteness a more solid status than a blind hypothesis in the face of a lack of deterministic explanation for quantum phenomena.

In~\cite{DBLP:conf/birthday/AbbottCS15} we used our model to prove that value indefiniteness can indeed be used to explain quantum unpredictability.
Specifically, we showed that
%
	\emph{If $E$ is an experiment measuring a quantum value indefinite projection observable, then the outcome of a single trial of $E$ is (simply) unpredictable.}

Although value indefiniteness guarantees unpredictability, it relies largely on, and is thus relative to, the Kochen-Specker theorem and its hypotheses~\cite{kochen1,pitowsky:218,2012-incomput-proofsCJ}, which only holds for systems in three or more dimensional Hilbert space.
It is thus useful to know if any other quantum phenomena can be used to certify unpredictability that would be present in two-dimensional systems or in the absence of other Kochen-Specker hypotheses, and if so, what degree of unpredictability is guaranteed.

\subsection{Complementarity}

The quantum phenomena of complementarity has also been linked to unpredictability and, contrary to the value indefiniteness pinpointed by the Kochen-Specker theorem, is present in all quantum systems.
By itself quantum complementarity is not \emph{a priori} incompatible with value definiteness (there exist automaton and generalised urn models featuring complementarity but not value indefiniteness~\cite{wright,svozil-2001-eua}) and hence constitutes a weaker hypothesis, even though it is sometimes taken as ``evidence'' when arguing that value indefiniteness is present in all quantum systems.

It is therefore of interest to see if complementarity alone can guarantee some degree of unpredictability, and is an ideal example to apply our model to.
This interest is not only theoretical, but also practical as some current quantum random generators~\cite{stefanov-2000} operate in two-dimensional Hilbert space where the Kochen-Specker theorem cannot be used to certify value indefiniteness, and would hence seem to (implicitly) rely on complementarity for certification.

\subsubsection{Quantum complementarity}

Let us first discuss briefly the notion of quantum complementarity, before we proceed to an analysis of its predictability.

The principle of complementarity was originally formulated and promoted by Pauli~\cite{pauli:58}.
It is indeed more of a general principle rather than a formal statement about quantum mechanics, and states that it is impossible to simultaneously measure formally non-commuting observables, and for this reason commutativity is nowadays often synonymous with co-measurability.
It is often discussed in the context of the position and momentum observables, but it is equally applicable to any other non-commuting observables such as spin operators corresponding to different directions, such as $S_x$ and $S_y$, which operate in two-dimensional Hilbert space.

Given a pair of such ``complementary'' observables and a spin-$\frac{1}{2}$ particle, measuring one observable alters the state of the particle so that the measurement of the other observable can no longer be performed on the original state.
Such complementarity is closely related to Heisenberg's original uncertainty principle~\cite{Heisenberg-27}, which postulated that any measurement arrangement for an observable necessarily introduced uncertainty into the value of any complementary observable.
For example, an apparatus used to measure the position of a particle, would necessarily introduce uncertainty in the knowledge of the momentum of said particle. 
This principle and supposed proofs of it have been the subject of longstanding (and ongoing) debate~\cite{Busch:2014lr,Cowen:2013fe,Rozema:2012hb}.

More precise are the formal uncertainty relations due to Robertson~\cite{Robertson:1929ee}---confusingly also often referred to as Heisenberg's uncertainty principle---which state that the standard deviations of the position and momentum observables satisfy $\sigma_x\sigma_p \ge \hbar/2$, and give a more general form for any non-commuting observables $A$ and $B$.
However, this mathematically only places constraints on the variance of repeated measurements of such observables, and does not formally imply that such observables cannot be co-measured, let alone have co-existing definite values, as is regularly claimed~\cite[Ch. 3]{Popper:1992ty}.

Nonetheless, complementarity is usually taken to mean the stronger statement that it is impossible to simultaneously measure such pairs of observables, and that such measurement of one will result in a loss of information relating to the non-measured observable following the measurement.
We will take this as our basis in formalising complementarity, but we do not claim that such a loss of information need be more than epistemic;
to deduce more from the uncertainty relations one has to assume quantum indeterminism---that is, value indefiniteness.

\subsubsection{Complementarity and value definiteness: a toy configuration}

In order to illustrate that complementarity is not incompatible with value definiteness we  briefly consider an example of a toy-model of a system that is value definite but exhibits complementarity.
This model was outlined in~\cite{svozil-2001-eua} and concerns a system modelled as an automaton; a different, but equivalent, generalised urn-type model is described in~\cite{wright}.

Although this example is just a toy model and does not correspond to a complete quantum system, it represents well many aspects of quantum logic, and serves to show that complementarity itself is not incompatible with value definiteness.

The system is modelled as a \emph{Mealy automaton} $\mathcal{A}=(S,I,O,\delta,W)$ where $S$ is the set of states, $I$ and $O$ the input and output alphabets, respectively, $\delta : S\times I \to S$ the transition function and $W: S \times I \to O$ the output function.
If one is uncomfortable thinking of a system as an automaton, one can consider the system as a black-box, whose internal workings as an automaton are hidden.
The state of the system thus corresponds to the state $s$ of the automaton, and each input character $a\in I$ corresponds to a measurement, the output of which is $W(s,a)$ and the state of the automaton changes to $s'=\delta(s,a)$.
To give a stronger correspondence to the quantum situation, we demand that repeated measurements of the same character $a\in I$ (i.e. observable) gives the same output: for all $s\in S$ $W(s,a)=W(\delta(s,a),a)$.
The system is clearly value definite, since the output of a measurement is defined prior to any measurement being made.

However, if we have two ``measurements'' $a,b\in I$ such that $W(s,a)\neq W(\delta(s,b),a)$ then the system behaves contextually;
$a$ and $b$ do not commute.
Measuring $b$ changes the state of the system from $s$ to $s'=\delta(s,b)$, and we lose the ability to know $W(s,a)$.

\subsection{Complementarity and unpredictability}

Complementarity tends to be more of a general principle than a formal statement, hence in order to investigate mathematically the degree of unpredictability that complementarity entails we need to give complementarity a solid formalism.
While several approaches are perhaps possible, following our previous discussion we choose a fairly strong form of complementarity and consider it not as an absolute impossibility to simultaneously know the values of non-commuting observables, but rather as a restriction on our current set of extractors---i.e.\ using standard quantum measurements and other techniques we currently have access to.

Formally, we say the set of extractors $\Xi$ is \emph{restricted by complementarity} if, for any two incompatible quantum observables $A,B$ (i.e., $[A,B]\neq 0$), there does not exist an extractor $\xi\in\Xi$ and a computable function $f$ such that, whenever the value $v(A)$ of the observable $A$ is known\footnote{We assume for simplicity that the observables $A$ and $B$ have discrete spectra (as for bounded systems), that is, the eigenvalues are isolated points, and hence the values $v(A)$ and $v(B)$ can be uniquely determined by measurement.
	Furthermore, since the choice of units is arbitrary (e.g., we can choose $\hbar=1$) one can generally assume that $v(A)$ and $v(B)$ are rational-valued, and hence can be known `exactly'.
	Even if this were not the case, a finite approximation of $v(A)$ is sufficient to uniquely identify it, and thus is enough here.

	For continuous observables it is obviously impossible to identify precisely $v(A)$ or $v(B)$.
	Such systems are generally idealisations, but one can still handle this case by considering observables $A'$ and $B'$ that measure $A$ and $B$ to some fixed accuracy.
	Protection by complementarity may depend on this accuracy. For example, for position and momentum, one expects complementarity to apply only when the product of accuracies in position and momentum is less than $\hbar/2$ according to the uncertainty relations.},
then for all $\lambda$, $f(\xi(\lambda))=v(B)$.

This states that, if we know $v(A)$ we have no way of extracting, directly or indirectly, the value $v(B)$ without altering the system.
We stress that this doesn't imply that $A$ and $B$ cannot simultaneously have definite values, simply that we cannot \emph{know} both at once.

Let us consider an experiment $E_C$ that prepares a system in an arbitrary pure state $\ket{\psi}$, thus giving $v(P_\psi)=1$ for the projection observable $P_\psi=\oprod{\psi}{\psi}$, before performing a projective measurement onto a state $\ket{\phi}$ with $0 < \iprod{\psi}{\phi} < 1$ (thus $[P_\psi,P_\phi]\neq 0$) and outputting the resulting bit.

It is not difficult to see that  this experiment is unpredictable relative to an agent whose predicting power is restricted by complementarity.
More formally, if a set of extractors $\Xi$ is restricted by complementarity, then the experiment $E_C$ described above is unpredictable for $\Xi$.
For otherwise, there would exist an extractor $\xi\in\Xi$ and a computable predictor $P_{E_C}$ such that, under any repetition procedure giving parameters $\lambda_1,\lambda_2,\dots$ we have $P_{E_C}(\xi(\lambda_i))=x_i$ for all $i$, where $x_i$ is the outcome of the $i$th iteration/trial.
But if we take $f=P_{E_C}$, then the pair $(\xi,f)$ contradicts the restriction by complementarity, and hence $E_C$ is unpredictable for $\Xi$.

It is important to note that this result holds regardless of whether the observables measured are value definite or not, although the value definite case is of more interest. 
Indeed, if the observables are value indefinite then we are guaranteed unpredictability without assuming restriction by complementarity, and hence we gain little extra by considering this situation.

As a concrete example, consider the preparation of a spin-$\frac{1}{2}$ particle, for instance an electron, prepared by in a $S_z=+\hbar/2$ state before measuring the complementary observable $2S_x/\hbar$ producing an outcome in $\{-1,+1\}.$
This could, for example, be implemented by a pair of orthogonally aligned Stern-Gerlach devices.
Next let us assume that the system is indeed value definite.
The preparation step means that, prior to the trial of the experiment being performed, $v(S_z)$ is known, and by assumption $v(S_x)$ exists (i.e.,\ is value definite) and is thus ``contained'' in the parameter $\lambda$.
The assumption that $\Xi$ is restricted by complementarity means that there is no extractor $\xi\in\Xi$ able  to be used by a predictor $P_E$ giving $P_E(\xi(\lambda_i))=2v(S_x)/\hbar=x_i$, thus giving unpredictability for $\Xi$.

As we noted at the start of the section, this is a fairly strong notion of complementarity (although not the strongest possible).
A weaker option would be to consider only that we cannot directly extract the definite values: that is, there is no $\xi\in\Xi$ such that $\xi(\lambda)=v(S_x)$, for all $\lambda$.
However, this does not rule out the possibility that there are other extractors allowing us to indirectly measure the definite values (unless we take the strong step of assuming $\Xi$ is closed under composition with computable functions, for example).
This weaker notion of complementarity would thus seem insufficient to derive unpredictability for $\Xi$, although it would not show predictability either.
We would thus, at least for the moment, be left unsure about the unpredictability of measurements limited by this weak notion of complementarity.

\section{Unpredictability, computability and complementarity}

In an effort to try and understand exactly how random quantum randomness---the randomness generated by measuring unpredictable quantum observables---actually is, we showed in~\cite{2012-incomput-proofsCJ} that quantum value indefiniteness leads to a strong form of incomputability.\footnote{Technically: A sequence $x_1x_2\dots$ is bi-immune if it contains no computable subsequence, that is, no computable function can compute exactly the values of more than finitely many bits of the sequence.}
Since this type of incomputability represents a notion of purely algorithmic unpredictability~\cite{DBLP:conf/birthday/AbbottCS15}, one may be tempted to think that this is a result not so much of quantum value indefiniteness, but rather of quantum unpredictability.

In~\cite{DBLP:conf/birthday/AbbottCS15}, however, we showed that this is not the case: there are  unpredictable experiments capable of producing both computable and strongly incomputable sequences when repeated \emph{ad infinitum}.
It is thus \emph{a fortiori} true that the same is true for relativised unpredictability, and there is no immediate guarantee that measurements of complementary observables must lead to incomputable sequences as is the case with value indefiniteness.

\subsection{Incomputability and complementarity}
\label{sec:ic}

Even though the (relativised) unpredictability associated with complementary quantum observables cannot guarantee incomputability, one may ask whether this complementarity may, with reasonable physical assumptions, lead directly to incomputability, much as value indefiniteness does.

Here we show this not to be true in the strongest possible way.
Specifically, we will show how an, admittedly toy, (value definite) system exhibiting complementarity (and thus unpredictable relative for extractors limited by the complementarity principle) can produce computable sequences when repeated.

Consider an experiment $E_M$ involving the prediction of the outcome of measurements on an (unknown) Mealy automaton $M=(Q,\Sigma,\Theta,\delta,\omega)$, which we can idealise as a black box, with $\{x,z\}\in \Sigma$ characters in the input alphabet, output alphabet $\Theta=\{0,1\}$ and satisfying the condition that $x$ and $z$ are complementary:
that is, for all $q\in Q$ we have $\omega(q,z)\neq \omega(\delta(q,x),z)$ and $\omega(q,x)\neq \omega(\delta(q,z),x)$.
This automaton is deliberately specified to resemble measurements on a qubit.
This  very abstract model can be viewed as a toy interpretation of a two-dimensional value definite quantum system, where the outcome of measurements are determined by some unknown, hidden Mealy automaton.
Since the Kochen-Specker theorem does not apply to two-dimensional systems, this value definite toy model poses no direct contradiction with quantum mechanics~\cite{kochen1}, even if it is not intended to be particularly realistic.
We complete the specification of $E_M$ by considering a trial of $E_M$ to be the output on the string $xz$, that is, if the automaton is initially in the state $q$, the output is $\omega(\delta(q,x),z)$, and the final state is $\delta(\delta(q,x),z)$.
This is a clear analogy to the preparation and measurement of a qubit using complementary observables, of the type discussed earlier.

Let us show that $E_M$ is unpredictable for a set $\Xi_C$ of extractors that expresses the restriction by complementarity present in Mealy automata.
In particular, let us  consider the set $\Xi_C$ that, in analogy to the restriction by complementarity of two quantum observables defined earlier, is restricted by an analogue of complementarity for the inputs $x,z \in \Sigma$ in the following sense:
\emph{there is no extractor $\xi\in\Xi_C$ and computable function $f$ such that, if $\lambda_M$ is the state of a system with Mealy automaton $M$ in a state $q$ such that $\delta(q,x)=q$ (or $\delta(q,z)=q$, that is, in an ``eigenstate''  $x$ or $z$), then $f(\xi(\lambda_M))=\omega(q,z)$ (or $f(\xi(\lambda_M))=\omega(q,x)$).}
That is, if $M$ is in an ``eigenstate'' of $x$, we cannot extract the output of the input $z$ (and similarly for $z$ and $x$ interchanged).

Let us assume for the sake of contradiction that $E_M$ is predictable for $\Xi_C$: that is, there is a predictor $P_{E_M}$ and an extractor $\xi\in\Xi_C$ such that $E_M$ is predictable for $\Xi_C$.
Thus, from the definition of predictability, the pair $(P_{E_M},\xi)$ must provide infinitely many correct predictions when repeated with the following iteration procedure (in analogy to preparing in an $x$ eigenstate): the black box containing $M$ is prepared by inputting ``$x$'', and then the experiment is run and the output recorded.
The next repetition is performed on the same system, preparing the box once again by inputting ``$x$'' and performing the experiment.
Thus, from the definition of Mealy automata, for each repetition $i$ the automaton $M$ is in a state $q_i$ such that $\delta(q_i,x)=q_i$ before the $i$th trial is performed.
Thus, the output of the $i$th trial of $E_M$ is precisely $\omega(\delta(q_i,x),z)=\omega(q_i,z)$, and for each trial we have $P_{E_M}(\xi(\lambda_i))=\omega(q_i,z)$, but since $P_{E_M}$ is a computable function this predictor/extractor pair contradicts the restriction by this form of complementarity of $\Xi_C$, and hence we conclude that $E_M$ is unpredictable for $\Xi_C$.

The main question is thus the (in)computability of sequences produced by the concatenation of outputs from infinite repetitions of $E_M$.
The experiment can be repeated under many different repetition scenarios, but the simplest is by  performing the experiment again on the same black box (and thus with the same automaton) with the final state of $M$ becoming the initial state for the next repetition\footnote{Recall that the internal state of $M$ is hidden and not part of the description of $E_M$, so there is no requirement that it be reset for each repetition.}.
In this case,  the sequence produced is computable---even cyclic---as a result of the automaton $M$ used.
Thus, even if this is not the case under all repetition scenarios, we cannot guarantee that the sequence produced is incomputable, even though $E_M$ is unpredictable for $\Xi_C$.

We note that one could easily consider slightly more complicated scenarios where the outcomes are controlled not by a Mealy automaton, but an arbitrary computable---or even, in principle, incomputable---function; complementarity is agnostic with respect to the computability of the output of such an experiment.
Such a computable sequence may be  obviously computable---e.g. $000\dots$, but it could equally  be something far less obvious, such as the digits in the binary expansion of $\pi$ at prime indices, e,g, $\pi_2\pi_3\pi_5\pi_7\pi_{11}\dots$.
Hence, this scenario cannot be easily ruled out empirically, regardless of the computability, that is,  low complexity, of the resulting sequences.
Further emphasising this, we note that computable sequences can also be Borel-normal, as in Champernowne's constant or (as conjectured) $\pi$, and thus satisfy many statistical properties one would expect of random sequences.

Our point was not to propose this as a realistic physical model---although it perhaps cannot be dismissed so easily---but to illustrate a conceptual possibility.
Value indefiniteness rules this computability out, but complementarity fails to do the same in spite of its intuitive interpretation as a form of quantum uncertainty.
At best it can be seen as an epistemic uncertainty, as it at least poses a physical barrier to the knowledge of any definite values.
The fact that complementarity cannot guarantee incomputability is in agreement with the fact that value definite, \emph{contextual} models of quantum mechanics are perfectly possible~\cite{DBLP:conf/birthday/AbbottCS15,Bohm52}; such models need not contradict any principle of complementarity, and can be computable or incomputable.

\section{Summary}

In this paper, following on from previous work in~\cite{DBLP:conf/birthday/AbbottCS15}, we developed a revised and more nuanced formal model of (un)predictability for physical systems.
By considering prediction agents with access to restricted sets of extractors with which to obtain information for prediction, this model allows various intermediate degrees of prediction to be formalised.

Although models of prediction such as this can be applied to arbitrary physical systems, we have discussed in detail their utility in helping to understand quantum unpredictability, which plays a key role in quantum information and cryptography.

We showed that, unlike measurements certified by value indefiniteness, those certified by complementarity alone are not necessarily simply unpredictable: \emph{they are unpredictable relative to the ability of the predicting agent to access the values of complementarity observables}---a more epistemic, relativised notion of predictability.
This is a general result about complementarity, not specifically in quantum mechanics, and certification by complementarity and value indefiniteness need not be mutually exclusive.
Indeed, in dimension three and higher Hilbert space, relative to the assumptions of the Kochen-Specker theorem~\cite{2012-incomput-proofsCJ} one has certification by both properties, value indefiniteness thus providing the stronger certification.
However, our results are of more importance for two-dimensional systems, since although quantum complementarity is present, this does not necessarily lead to value indefiniteness.
While one may postulate value indefiniteness in such cases as well, this constitutes an extra physical assumption, a fact which should not be forgotten~\cite{DBLP:conf/birthday/AbbottCS15}.
In assessing the randomness of quantum mechanics, one thus needs to take carefully into account all physical assumptions contributing towards the conclusions that one reaches.

The fact that quantum complementarity provides a weaker certification than value indefiniteness is emphasised by our final result, showing that complementarity is compatible with the production of computable sequences of bits, something not true for value indefiniteness.
Thus, quantum value indefiniteness and the Kochen-Specker theorem appear, for now, essential in certifying the unpredictability and incomputability of quantum randomness.

\begin{acknowledgments}
This work was supported in part by Marie Curie FP7-PEOPLE-2010-IRSES Grant RANPHYS.
\end{acknowledgments}

\if01
\authorcontributions{Author Contributions}

The authors all contributed equally to this paper.


\conflictofinterests{Conflicts of Interest}

The authors declare no conflict of interest.

\fi


\bibliography{svozil,alastairsRefs}

\end{document}